\documentclass[sigconf]{acmart}

\AtBeginDocument{%
  }

\setcopyright{acmlicensed}
\copyrightyear{2024}
\acmYear{2024}
\acmDOI{10.1145/3697789.3697794}

\acmConference[ICGJ '24]{the 8th International Conference on Game Jams,
Hackathons and Game Creation Events}{October 11, 2024}{Copenhagen, Denmark}
\acmISBN{979-8-4007-1779-6/24/10}

\usepackage{xspace}
\usepackage{lscape}

\newcommand{\SaoPaulo}[0]{State of São Paulo\xspace}
\newcommand{\UniversityOfSaoPaulo}[0]{University of São Paulo\xspace}
\newcommand{\USP}[0]{USP\xspace}
\newcommand{\InstituteOfMathematicsAndStatistics}[0]{Institute of Mathematics and Statistics\xspace}
\newcommand{\IME}[0]{IME\xspace}
\newcommand{\ICMC}[0]{ICMC\xspace}
\newcommand{\EACH}[0]{EACH\xspace}
\newcommand{\Brazil}[0]{Brazil\xspace}
\newcommand{\Brazilian}[0]{Brazilian\xspace}
\newcommand{\CodeLab}[0]{CodeLab\xspace}
\newcommand{\HackathonIME}[0]{HackathonIME\xspace}
\newcommand{\HackathonUSP}[0]{HackathonUSP\xspace}
\newcommand{\InterHack}[0]{InterHack\xspace}
\newcommand{\HackFools}[0]{HackFools\xspace}
\newcommand{\SheHacks}[0]{SheHacks\xspace}
\newcommand{\DeepHack}[0]{DeepHack\xspace}
\newcommand{\HackMobilidade}[0]{HackMobilidade\xspace}
\newcommand{\FreeSoftwareCompetenceCenter}[0]{Free Software Competence Center\xspace}
\newcommand{\CCSL}[0]{CCSL\xspace}
\newcommand{\Reais}[0]{Brazilian Reais\xspace}

\begin{document}


\title{
  The Journey of \CodeLab: How University Hackathons
  Built a Community of Engaged Students%
}



\author{Renato Cordeiro Ferreira}
\orcid{0000-0001-7296-7091}
\affiliation{%
  \institution{Institute of Mathematics and Statistics, University of São Paulo}
  \city{São Paulo}
  \country{Brazil}
}
\email{renatocf@ime.usp.br}

\author{Renata Santos Miranda}
\orcid{0009-0006-8927-1756}
\affiliation{%
  \institution{Institute of Mathematics and Statistics, University of São Paulo}
  \city{São Paulo}
  \country{Brazil}}
\email{renata.miranda@usp.br}

\author{Alfredo Goldman}
\orcid{0000-0001-5746-4154}
\affiliation{%
  \institution{Institute of Mathematics and Statistics, University of São Paulo}
  \city{São Paulo}
  \country{Brazil}}
\email{gold@ime.usp.br}

\renewcommand{\shortauthors}{Ferreira et al.}


\begin{abstract}
  This paper presents the journey of \CodeLab: a student-organized
  initiative from the \UniversityOfSaoPaulo that has grown thanks to
  university hackathons.
  It summarizes patterns, challenges, and lessons learned over
  15 competitions organized by the group from 2015 to 2020.
  By describing these experiences, this report aims to help \CodeLab
  to resume its events after the \mbox{COVID-19} pandemic, and foster
  similar initiatives around the world.
\end{abstract}

\ccsdesc[500]{Human-centered computing~Collaborative and social computing}

\keywords{University Hackathons, CodeLab, Community Building}


\begin{teaserfigure}
  \centering
  \includegraphics[width=0.97\textwidth]{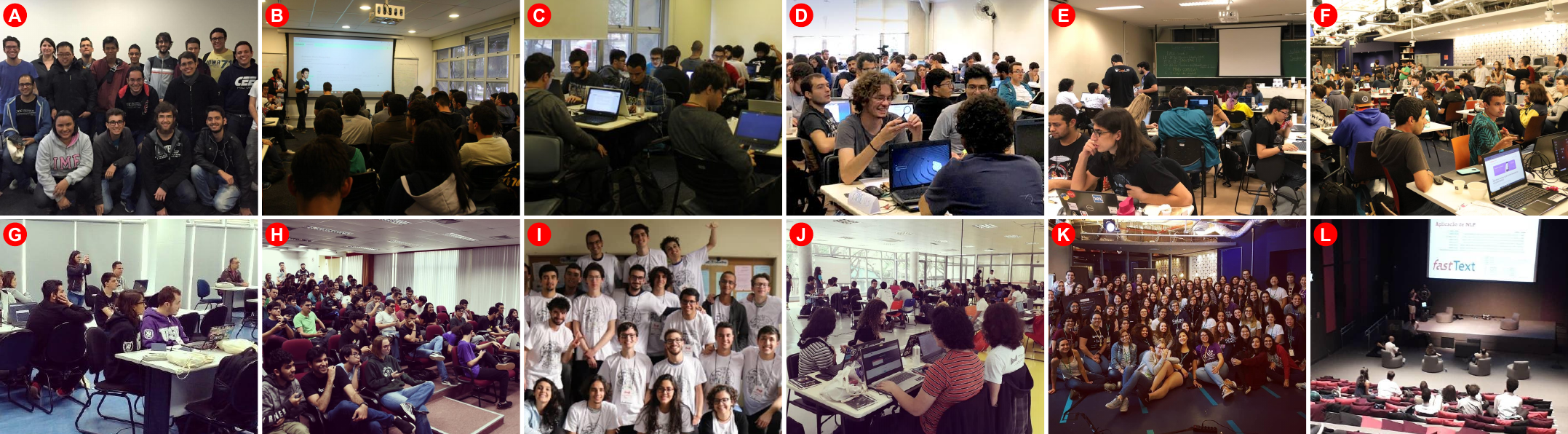}
  \captionsetup{skip=0.5em}
  \caption{%
    \CodeLab Events:
    HackathonIME$^{A}$,
    HackathonUSP$^{B-E}$,
    HackFools$^{F}$,
    InterHack$^{G-J}$,
    SheHacks$^{K}$,
    DeepHack$^{L}$.
  }
  \Description{
    Composition containing 12 pictures of different events organized
    by CodeLab throughout its history:
      A - HackathonIME 2015,
      B - HackathonUSP 2017.1,
      C - HackathonUSP 2017.2,
      D - HackathonUSP 2018.1,
      E - HackathonUSP 2018.2,
      F - HackFools 2019,
      G - InterHack Selective at ICMC,
      H - InterHack Selective at EACH,
      I - InterHack Selective at IME,
      J - InterHack Final,
      K - SheHacks 2019,
      L - DeepHack 2019.
  }
  \label{fig:teaser}
  \vspace{0.8em}
\end{teaserfigure}


\received{28 June 2024}


\maketitle


\section{Introduction}\label{sec:introduction}

\CodeLab is a student-organized initiative conducted at the
\UniversityOfSaoPaulo (\USP), \Brazil. Its mission is to foster technological
innovation, helping students of Computer Science and Information Systems
to become professional Software Engineers.
Group members organize programming courses and study groups, where they
  teach and learn new technologies,
  discuss development practices from industry,
  and study research from academia.

\CodeLab employs hackathon-like events to show how software development
can solve real-world problems.
The group has been organizing these events since 2015, attracting students
from all campuses of the university%
\footnote{\USP is the largest public university in \Brazil, ranked by 
\href{https://www.timeshighereducation.com/world-university-rankings/university-sao-paulo}{Times Higher Education in 2023 among the 100 best in the world}.
It has campuses in 8 cities across the \SaoPaulo,
four of which offer bachelors in Computer Science and/or Information Systems.}.
As a side effect, the competitions drive the interest of participants in
joining \CodeLab.
This created a virtuous cycle, making the growth of \CodeLab's
community intertwined with the experience of organizing these events.

The aim of this paper is to report about 15 competitions organized by
\CodeLab from 2015 to 2020. It highlights patterns, challenges, and
lessons learned while organizing these events, which led to the creation
of a successful multi-campus community composed of up to $90$ students
at its peak.

Unfortunately, the COVID-19 pandemic impacted \CodeLab's institutional knowledge
of organizing hackathons. This paper is also an effort to document these
experiences for future members, while providing insights for similar
initiatives around the world.

  Section~\ref{sec:hackathons} outlines the types of event organized by \CodeLab,
  presenting data about them. Sections~\ref{sec:HackathonUSP} to~\ref{sec:DeepHack}
  explore the competitions organized by the group, describing their goal, history,
  and impact. Finally, section~\ref{sec:final_remarks} presents the final remarks.


\section{\CodeLab's University Hackathons}\label{sec:hackathons}

From 2015 to 2020, \CodeLab members organized 15 competitions for university
students. \autoref{tab:hackathon_data} summarizes all data collected about
these events in years of activity of the group.

For each event, the goal was to create a prototype of software or hardware
related to a predefined theme. Participants were divided into self-organized
teams of 3--5 people. Submissions were evaluated by 3--5 judges from industry,
academia, and organization. Prizes and honorable mentions were awarded according
to pre-established criteria. This format was inspired by hackathons organized by
Meta (formerly Facebook), one of the first companies to bring these competitions
in \Brazil~\cite{HackFacebook2012}. Most events happened during weekends to
avoid conflicts with classes.

After years of experience, \CodeLab members nicknamed four patterns of
competitions (also used for marketing purposes). These names are used
along the paper to differentiate each type of event:
\begin{itemize}
  \item \emph{Hackathon}, a sleepover competition up to 30h long;
  \item \emph{Hackday}, a continuous competition up to 10h long;
  \item \emph{Hackfest}, a many-days intermittent competition;
  \item \emph{Hack Championship}, a multi-stage competition composed of
        at least two events of the previous types.
\end{itemize}

\begin{table}[bh]
  \centering
  \includegraphics[width=\linewidth]{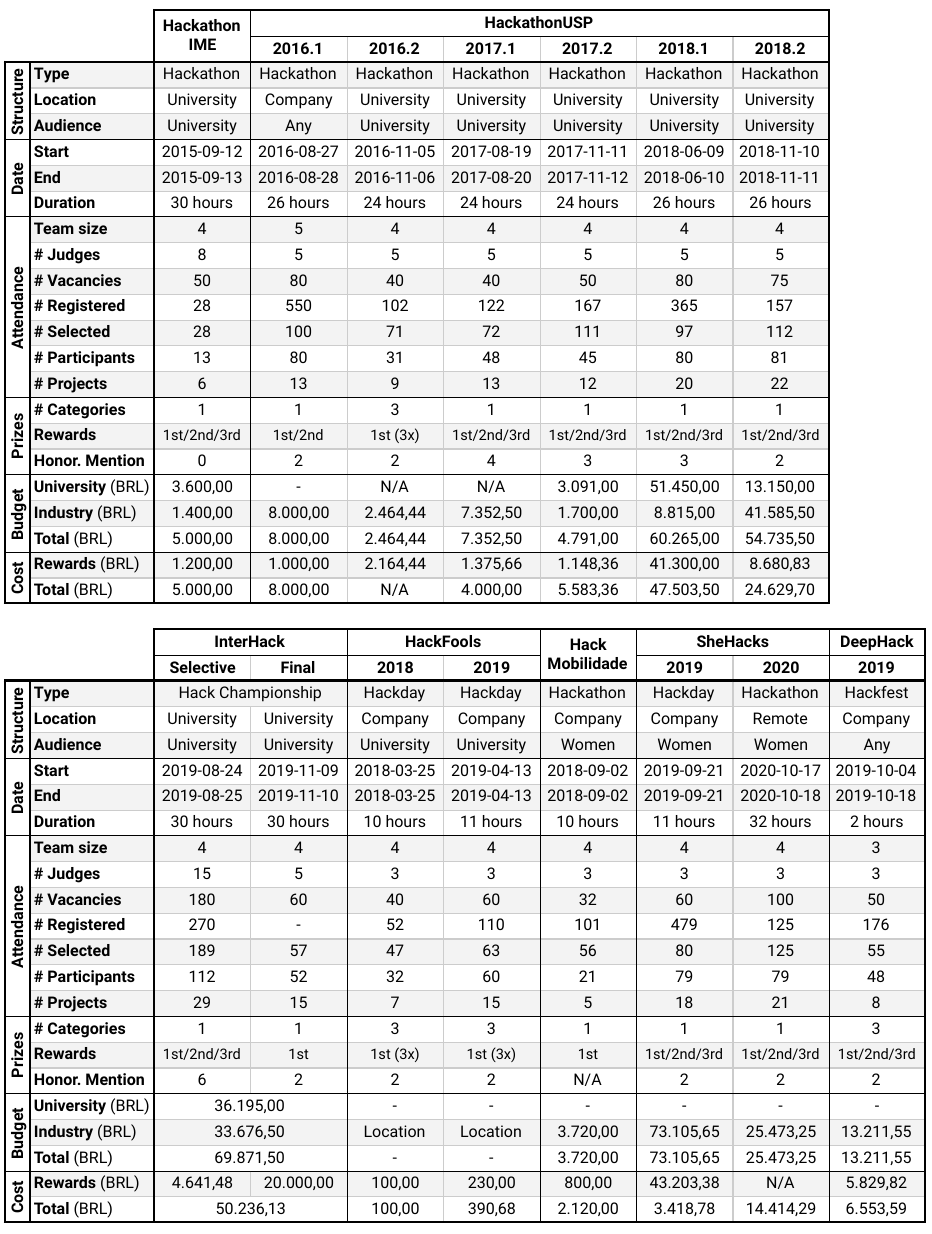}
  \vspace{-0.8em}%
  \caption{%
    Data about events organized by \CodeLab from 2015 to 2020.
    Not available (N/A) data was not managed by the group.
    All monetary values are in \Reais (BRL).
  }
  \label{tab:hackathon_data}
\end{table}



\section[\HackathonIME]{\HackathonIME~-- The First Experience}
\label{sec:HackathonIME}

Since 2012, hackathons have been common in \Brazil, organized both by
companies \cite{HackFacebook2012} and the government \cite{HackBus2013}.
Many competitions, in particular by Meta (formerly Facebook)
\cite{HackFacebook2012}, targeted students from top universities
for internship and hiring. As a consequence, many undergraduates from the
\UniversityOfSaoPaulo (\USP) had the opportunity of joining these events by
their volition.

In 2015, a group of senior students of the \InstituteOfMathematicsAndStatistics
(\IME) organized \HackathonIME: the first competition of its kind at the
\USP. Some organizers -- including the first author of this
paper -- had previous experience participating in the aforementioned company
hackathons. However, it was difficult to stimulate junior students to join
these events, as their programming experience was superficial.

The goal of \HackathonIME was to create an event friendly to first-time
hackathon participants. \HackathonIME had 50 spots available, and 28
registrations. All were selected, resulting in 13 participants delivering three
projects. Unfortunately, the number of registrations was smaller than intended.
However, as the first attempt of organizing a hackathon, it was deemed a
success.

The experience with \HackathonIME stimulated the same group of senior students
to create \CodeLab: a self-organized dedicated initiative where they could study
more software development and continue organizing hackathons. The group started
with a few members supervised by a professor. They organized the first activities
-- development courses, workshops, and study groups -- for other undergraduates
at \IME.


\section[\HackathonUSP]{\HackathonUSP~-- Engagement and Innovation}
\label{sec:HackathonUSP}

In August 2016, exploring a partnership with a growing \Brazilian fintech,
\CodeLab organized its first hackathon as a group. The event was named
\HackathonUSP, a rebranding aimed to reach the broader community.
Thanks to the sponsorship, the event was hosted off-campus, at the partner
company's headquarters. This allowed opening the competition for students
enrolled in any \Brazilian public universities, not only the
\UniversityOfSaoPaulo.

\paragraph{\HackathonUSP 2016.1}
Initially, the organizers planned 60 spots available. However, to their
surprise, the competition achieved 550 registrations in three days.
Registration-wise, this made the first \HackathonUSP one of the most successful
events ever made by \CodeLab. Consequently, the organizers increased the
available spots to 80. Nevertheless, since the registrations were free, 100
people were selected to account for no-shows. In the end, the event had a total
of 80 participants with 13 projects delivered. To accommodate all participants
in the venue, teams could have up to five members, one of the few deviations
from the usual 4-people teams.

The success of this event helped to debut the \CodeLab initiative
at the \UniversityOfSaoPaulo.

\noindentparagraph{\indent}
In 2016, the group made a partnership with the student-led entrepreneurship 
league\footnote{\href{https://www.uspempreende.org/}{NEU -- Núcleo de
Empreendedorismo da \USP}}. Together, both groups organized another five
editions of \HackathonUSP. All these events were sponsored by the Dean's
Research Office. Therefore, they were restricted to \USP's students and
were hosted on the university's main campus.

Every theme introduced challenges whose solutions could benefit \USP.
Moreover, they selected as they were approachable by students with different
levels of knowledge: 
\begin{itemize}
    \item 2016.2: Ethics, Transparency and Public Administration
    \item 2017.1: Improving the Scientific Production
    \item 2017.2: Smart Cities
    \item 2018.1: How Data Science Can Impact Students' Life
    \item 2018.2: University Financial Sustainability
\end{itemize}

\paragraph{\HackathonUSP 2016.2}
In November 2016, the second edition had 40 spots available and 102
registrations. 71 people were selected, resulting in 31 participants
delivering 9 projects. This event had a large dropout rate ($56\%$).
Choosing a date at the end of the term, as well as scheduling two
competitions so close (with three months of interval) likely
impacted the participation.

\paragraph{\HackathonUSP 2017.1}
\CodeLab planned to grow the event alongside its other activities.
In August 2017, the competition had 40 spots available and 122 registrations.
72 people were selected, resulting in 48 participants delivering 13 projects.
Its dropout rates ($33\%$) were compatible with what was observed in the
2016.1 edition.

\paragraph{\HackathonUSP 2017.2}
Initially, the group did not plan to make a second event that same year,
but it happened thanks to a partnership with the \FreeSoftwareCompetenceCenter
(\CCSL) at \USP.
In November 2017, the competition had 50 spots available and 167 registrations.
111 people were selected, resulting in 45 participants delivering 12 projects.
Unfortunately, once again, the second event of the year had a large dropout
rate ($59\%$), which confirmed the bias against organizing two events in the
same term. 


\paragraph{\HackathonUSP 2018.1}
To avoid the large dropout rates observed previously, \CodeLab
started splitting the events at the end of each academic term.
In June 2018, the competition had 80 spots available and 365 registrations.
97 people were selected, resulting in 80 participants delivering 20 projects.
Although it was the end of the term, the dropout rate was stable ($33\%$).
Among the reasons, the prize -- a sponsored visit to Silicon Valley --
led to many registrations.

\paragraph{\HackathonUSP 2018.2}
In November 2018, the competition had 75 spots available and 157 registrations.
112 people selected, resulting in 81 participants delivering 22 projects.
Its theme was related to financial technology.
Both 2018 events were the largest, most successful editions of
\HackathonUSP. In particular, the 2018.2 edition had the second-largest
number of registrations. The impact achieved by the 2018.1 event influenced
the participation.

\noindentparagraph{\indent}
\HackathonUSP had a long-lasting impact on \CodeLab.
Students became excited about the activities organized by the group,
in particular, hackathon organization. The 2016 events established the group,
while the 2017 events helped it to expend from two to ten members.
Finally, the 2018 editions conceived two new local chapters of \CodeLab,
located in \ICMC and \EACH, where there are other Computer Science and
Information System courses. 


\section[\InterHack]{\InterHack~-- Reach and Impact}
\label{sec:InterHack}

The 2018 events showed that \USP still had a large, unattended demand for
hackathons. Therefore, \CodeLab wanted to grow its competitions, and train
the new members on how to organize them.

Unfortunately, \HackathonUSP as a single competition reached
saturation: available sites could no longer support the number of participants
($80$). Moreover, project presentations were taking longer, making it strenuous
for judges to choose prototypes during the event. This led \CodeLab to
organizing \InterHack: a hack championship composed of two phases. 


\paragraph{InterHack 2019 Selective}
In August 2019, \CodeLab organized the three hackathons simultaneously,
hosted by each chapter at their campus. The selective had 180 spots available
(60 for each event) and 270 registrations. 189 people were selected, resulting
in 112 participants delivering 29 projects across the events. The theme was
``how to improve the university management using technology''.

\paragraph{InterHack 2019 Final}
In November 2019, \CodeLab organized a final hackathon, which occurred in
the new \USP Innovation Center. The final had 60 spots available (20 for each
selective). 57 people were selected among those who were awarded prizes and
honorable mentions on each selective. 52 participants delivered 15 projects.
Its theme was revealed only at the beginning of the competition:
``how to improve education via technology''.

\noindentparagraph{\indent}
To the best of our knowledge, \InterHack was the biggest university
hackathon-based competition made in \Brazil at the time. Sponsorship-wise,
it was \CodeLab's biggest event, with support from the university and
private companies. It was also the most logistically challenging,
since the selective required planning and executing three hackathons
on different locations simultaneously.

\InterHack became a milestone of maturity for \CodeLab.
The championship led to partnerships with many companies (startups
to big tech) and strong support from the university (by the research,
the innovation, and the undergraduate Dean's Offices). Thanks to its
popularity, the group reached its maximum number of active members ($90$),
who organized many activities: lectures, courses, workshops, study groups,
and hackathons.

Unfortunately, due to the COVID-19 pandemic, plans to organize another
\InterHack were successively canceled. \CodeLab lost institutional knowledge
about hackathons, since new members could neither participate nor learn to
organize these events.


\section[\HackFools]{\HackFools~-- Excitement and Integration}
\label{sec:HackFools}

With the evolution from \HackathonUSP to \InterHack,
\HackFools was created to become the first opportunity for junior students to
join a hackathon -- one of the original goals of \HackathonUSP.
Participants develop a funny -- although fundamentally useless -- application.
The goal was to introduce hackathons in a jokey, lighthearted way. Moreover,
to provide a safe environment, where students do not feel pressure,
similarly to a Coding Dojo~\cite{Dojo}.

\paragraph{\HackFools 2018}
In April 2018, senior \CodeLab members organized the event to train new members
on hackathon organization. This competition had 40 spots available and 52 registrations.
47 people were selected, resulting in 32 participants delivering 7 projects.

\paragraph{\HackFools 2019}
In April 2019, \HackFools also aimed to train and integrate after members from the
two new \CodeLab chapters, This competition had 60 spots available (20 for each campus)
and 110 registrations. 63 people were selected, resulting in 60 participants
delivering 15 projects.


\section[\SheHacks]{\SheHacks~-- Leadership and Diversity}
\label{sec:SheHacks}

In 2018, \CodeLab made its first attempt to organize a hackathon exclusively
for women. Since then, \SheHacks has become one of the most successful competitions organized by the group, and a great tool to recruit more female participants for \CodeLab.


\paragraph{HackMobilidade}
In September 2018, \HackMobilidade happened as a hackathon related to
active mobility. The event was exclusively for women. This decision, however,
happened by chance, supported by the female-led sponsors.
This competition had 32 spots available and 101 registrations. 56 people were
selected, resulting in 21 participants delivering 5 projects. The number of
registrations showed the potential and high interest for women-only events.


\paragraph{SheHacks 2019}
In August 2019, the first \SheHacks was organized based on the experiences of
\HackMobilidade. Besides the public restriction (students who self-identify as
women), it was fully organized by a female team. The hackday format was chosen
to consider the safety and comfort of the participants. For example, this
avoided their need to leave the event at night in a personal emergency.

This competition had 60 spots available and 479 registrations.
80 people were selected, resulting in 79 participants delivering 18 projects.
The number of registrations was the second largest in \CodeLab's history (eight
times the number of spots). Moreover, it had the lowest dropout rate from
among all competitions.

There are three hypotheses for this success. First, this event had the largest
amount of private sponsorship ever obtained by \CodeLab, since companies wanted
to promote diversity. Second, the sponsorship allowed the group to offer valuable
prizes. Lastly, the event happened in the former Meta company headquarter in
\Brazil, which increased the attention to the event.

\paragraph{SheHacks 2020}
In 2020, \SheHacks happened remotely due to the COVID-19 pandemic.
It was the first online competition organized by \CodeLab. Since the
event was remote, the organizer committee understood that participants
would have less focus at home. Therefore, the event was expanded to a
hackathon, so all participants could have more time. Most importantly,
this decision did not affect the safety and comfort as it could
have in an in-person event.

The competition had 100 spots available and 125 registrations. 125 people were
selected, resulting in 79 participants delivering 22 projects. The longer
format allowed a larger number of spots available since there was no physical
space constraint.

Even though the competition had good prizes, the number of registrations
was much smaller. The hypothesis is that, missing an attractive location,
fewer people got interested. Moreover, the dropout rate was bigger, since
participants could decide last-minute to leave the event. Nevertheless,
the event was deemed as very successful as \CodeLab's first remote
competition.

\noindentparagraph{\indent}
\SheHacks was responsible for the increase in the number of women at \CodeLab,
Since the event was organized only by women, it promoted leadership and
protagonism among female \CodeLab members.
Before \SheHacks, there was an average of 22\% of people who identified
as women in \CodeLab events. \DeepHack 2019, which happened a month after
\SheHacks 2019, had 36\%. This impact is attributed to how the group's
social networks started reaching a more diverse demographic.


\section[\DeepHack]{\DeepHack~-- Partnership and Contribution}
\label{sec:DeepHack}

\DeepHack is a hackfest about data science, machine learning and
artificial intelligence, whose goal is to promote partnerships with
public institutions. The longer-format event was due to the complexity
of working with large volumes of data, which requires more time to explore
and formulate ideas.

\paragraph{Previous Experiences}
\HackathonUSP 2018.1 and \HackMobilidade were the first experiences when
\CodeLab brought themes related to data science. They highlighted some
particularities in this subject. Unlike other events, projects usually
generated simple data analyses rather than proposing full products.
These results were greatly influenced by the duration of the events:
teams did not have enough time to explore the data and apply machine
learning techniques.




\paragraph{\DeepHack 2019}
In October 2019, the event was hosted as a hackfest. The competition had 50
spots available and 176 registrations. 55 people were selected, resulting in 48
participants delivering 8 projects. The event was the first to explore
partnerships with public institutions, which offered data and proposed
problems.

Overall, the experience with \DeepHack was successful. A longer timespan allowed
participants to propose more complete solutions, with in depth analyses that
could be transformed into products. Moreover, it opened space for a more
thorough pre-event, including lectures and workshops about the subject.

Unfortunately, after the COVID-19 pandemic, there were no more attempts to
organize another edition of \DeepHack, either remotely or in-person. Since
a hackfest demanded more commitment from organizers, it was more affected
by the loss of institutional knowledge in \CodeLab.


\section{Final Remarks}\label{sec:final_remarks}

This paper describes the hackathon-like events organized by \CodeLab, as it
became one of the largest student-led initiatives at \USP, spread on its many
campuses. It summarizes patterns, challenges, and lessons learned over 15
events between 2015 and 2020.

Although the CODIV-19 pandemic had a great toll on the group's institutional
knowledge, the experiences reported in this paper are helping \CodeLab to resume
its events. Based on them, as of 2024, the group is organizing new editions of
\HackathonUSP, \HackFools, and \SheHacks, so it may keep impacting \USP's
community.






\begin{acks}
  We would like to thank all members of \CodeLab, current and former.
  Special thanks to Leonardo Lana, João Daniel, Carolina Arenas, and
  Daniel Cordeiro, who helped recollect the experiences here.
  Finally, thanks to \USP, \IME, \ICMC, \EACH, and \CCSL for hosting the group.
\end{acks}


\bibliographystyle{ACM-Reference-Format}
\bibliography{codelab-refs}


\begin{thebibliography}{3}


\ifx \showCODEN    \undefined \def \showCODEN     #1{\unskip}     \fi
\ifx \showDOI      \undefined \def \showDOI       #1{#1}\fi
\ifx \showISBNx    \undefined \def \showISBNx     #1{\unskip}     \fi
\ifx \showISBNxiii \undefined \def \showISBNxiii  #1{\unskip}     \fi
\ifx \showISSN     \undefined \def \showISSN      #1{\unskip}     \fi
\ifx \showLCCN     \undefined \def \showLCCN      #1{\unskip}     \fi
\ifx \shownote     \undefined \def \shownote      #1{#1}          \fi
\ifx \showarticletitle \undefined \def \showarticletitle #1{#1}   \fi
\ifx \showURL      \undefined \def \showURL       {\relax}        \fi
\providecommand\bibfield[2]{#2}
\providecommand\bibinfo[2]{#2}
\providecommand\natexlab[1]{#1}
\providecommand\showeprint[2][]{arXiv:#2}

\bibitem[Agrela(2013)]%
        {HackBus2013}
\bibfield{author}{\bibinfo{person}{Lucas Agrela}.} \bibinfo{year}{2013}\natexlab{}.
\newblock \bibinfo{title}{'Hackatona do Ônibus' de São Paulo dá prêmios de até R\$ 8 mil | Exame}.
\newblock
\newblock
\urldef\tempurl%
\url{https://exame.com/tecnologia/hackatona-do-onibus-de-sao-paulo-da-premios-de-ate-r-8-mil/}
\showURL{%
\tempurl}
\newblock
\shownote{Visited at 2021-05-05}.


\bibitem[{O Globo}(2012)]%
        {HackFacebook2012}
\bibfield{author}{\bibinfo{person}{{O Globo}}.} \bibinfo{year}{2012}\natexlab{}.
\newblock \bibinfo{title}{São Paulo recebe o primeiro Facebook Hackathon Brasil - Jornal O Globo}.
\newblock
\newblock
\urldef\tempurl%
\url{https://oglobo.globo.com/economia/sao-paulo-recebe-primeiro-facebook-hackathon-brasil-4933830}
\showURL{%
\tempurl}
\newblock
\shownote{Visited at 2021-05-05}.


\bibitem[Sato et~al\mbox{.}(2008)]%
        {Dojo}
\bibfield{author}{\bibinfo{person}{Danilo~Toshiaki Sato}, \bibinfo{person}{Hugo Corbucci}, {and} \bibinfo{person}{Mariana~Vivian Bravo}.} \bibinfo{year}{2008}\natexlab{}.
\newblock \showarticletitle{Coding Dojo: An Environment for Learning and Sharing Agile Practices}. In \bibinfo{booktitle}{\emph{Agile 2008 Conference}}. \bibinfo{publisher}{IEEE}, \bibinfo{address}{Toronto, Canada}, \bibinfo{pages}{459--464}.
\newblock
\showISBNx{978-0-7695-3321-6}
\urldef\tempurl%
\url{https://doi.org/10.1109/Agile.2008.11}
\showDOI{\tempurl}


\end{thebibliography}

\end{document}